\documentstyle[11pt,aaspp4]{article}

\def\t0{\theta_{\circ}}

\def\be{\begin{equation}}
\def\en{\end{equation}}

\def\etal{et al.\ }
\def\msun{M_{\sun}}
\def\rsun{R_{\sun}}

\def\msunyr{M_{\sun} yr^{-1}}

\def\mdot{\dot{M}}

\def\curf{{\cal F}}

\begin{document}

\title
{The Structure and Emission of the Accretion Shock in T Tauri Stars II: the Ultraviolet Continuum Emission}
\author{ Erik Gullbring \altaffilmark{1}, Nuria Calvet \altaffilmark{2,3}, 
James Muzerolle\altaffilmark{2,4}, and
Lee Hartmann\altaffilmark{2}}
\altaffiltext{1}{Stockholm Observatory, SE-133 36 Saltsj\"obaden, Sweden;
Electronic mail: erik@astro.su.se, ncalvet@cfa.harvard.edu, lhartmann@cfa.harvard.edu, jmuzerolle@cfa.harvard.edu}
\altaffiltext{2}{Harvard-Smithsonian Center for Astrophysics, 60 Garden St., Cambridge, MA 02138, USA}
\altaffiltext{3}{Also Centro de Investigaciones de Astronomia, M\'erida, Venezuela}
\altaffiltext{4}{Five College Astronomy Department, University of Massachusetts}

\begin{abstract}

We compare accretion shock models with optical and ultraviolet spectra of
pre-main sequence stars to (1) make the first determinations of accretion rates
in intermediate mass T Tauri stars from continuum emission, and (2) 
derive improved estimates of accretion rates and extinctions for continuum
T Tauri stars.  Our method extends the shock models developed by Calvet
\& Gullbring to enable comparisons with optical and archival 
{\it International Ultraviolet Explorer} ultraviolet spectra.
We find good agreement between the observations and the model predictions,
supporting the basic model of magnetospheric accretion shocks as well
as previous determinations of accretion rates and interstellar reddening
for the low mass T Tauri stars. The accretion rates determined for the
intermediate-mass T Tauri stars agree well with values obtained
through other methods that use near-infrared hydrogen
line strengths (Muzerolle et al. 1998).

\end{abstract}

\keywords{Accretion, accretion disks, Stars: Circumstellar Matter,
Stars: Formation, Stars: Pre-Main Sequence}

\section{Introduction}

More than twenty five years ago Lynden-Bell \& Pringle (1974) pointed
out the importance of an accretion disk as the source of excess
emission of T Tauri stars (TTS): infrared emission from the outer disk, and
ultraviolet emission from the boundary layer where 
disk material in Keplerian rotation settles down on the slowly rotating stellar
surface.  When the existence of circumstellar disks around
pre-main sequence stars became observationally established, mainly
from infrared observations by {\it IRAS}, attempts were
made to model the excess emission
that veils the absorption lines at optical and ultraviolet (UV)
wavelengths by means of emission from boundary layer regions
(Kenyon \& Hartmann 1987; Bertout et al. 1988; Basri \& Bertout 1989).

In the last few years a different picture for loading the disk
material onto the surface of the star has
developed. In this scenario, the accretion disk is disrupted at a few
stellar radii by the stellar magnetic field, and the material falls
along the field lines at nearly free-fall velocity. The 
evidence for magnetospheric accretion is based mainly on three
observational properties: (1) the broad emission lines, often with
redshifted absorption components, are consistent with emission from
infalling gas channeled by the magnetosphere (Calvet \& Hartmann 1992;
Hartmann, Hewett, \& Calvet 1994; Muzerolle, Calvet \& Hartmann
1998; Muzerolle, Hartmann \& Calvet 1998a,b); (2) the near 
infrared colors and spectral energy distributions
of Classical T Tauri stars (CTTS) indicate emission from disks that
are truncated at a few stellar radii (Kenyon et al. 1994; Meyer,
Calvet, \& Hillenbrand 1997); (3) 
one of the three distinct types of variability
displayed by CTTS (Type II variations)
%the periodic modulation in the
%light curves of CTTS 
can be interpreted as emission from hot spots 
on the stellar surface, 
where infalling material impacts the star
(Herbst et al. 1994 [and references therein];
Gullbring et al. 1996) Thus,
magnetospheric accretion provides a self consistent picture to explain
the excess emission and many of the emission lines.

A key prediction of the magnetospheric picture is that a shock forms
close to the stellar surface, where much of the excess continuum and
line emission originates. Since shock models are used to derive
physical parameters from the observed excesses (cf. Gullbring et al. 1998,
GHBC), we need to test the
shock theory to make sure the interpretations are made correctly in
assuming magnetospheric infall.

In Calvet \& Gullbring (1998; hereafter Paper~I), we showed that the
spectral energy distribution of the excess emission in the optical
can be explained as a combination of optically thick
emission from the heated photosphere below the shock, dominating the
Paschen and Brackett continua, and optically thin emission from the
pre-shock and attenuated post-shock regions, becoming important
at wavelengths
shorter than the Balmer threshold. 
In this paper, we extend our analysis combining both
optical and ultraviolet data,
using {\it International Ultraviolet Explorer}
(IUE) spectra, to provide an additional test to the accretion shock
model in the wavelength region where shock emission should completely
dominate. We find that the accretion shock models can explain
both the excess flux that veils the absorption lines
in the optical spectra and the large ultraviolet continuum
fluxes in T Tauri stars. This result is in agreement
with that of Ardila and Basri (2000),
who have used 
our shock models to interpret the IUE data of T Tauri stars.
Ardila and Basri also conclude that our models can
explain the overall features of the ultraviolet spectra,
although more refinements are needed
to understand the details of the variability.   

The change of accretion rate with stellar age, as the pre-main
sequence stars evolve towards the main sequence, provides important
information on the evolution and physical processes of circumstellar
disks (Hartmann et al. 1998). Unfortunately, to
date no direct way exists to measure the accretion luminosity for
intermediate mass stars. The reason is that at optical wavelengths,
the intrinsic stellar luminosity for stars with spectral types
earlier than K5 dominates over the accretion emission, making it
difficult to measure the latter. However, at ultraviolet wavelengths the
stellar contribution drops significantly, and intermediate mass T Tauri
stars do show excess emission there. The lack of reliable models for
the UV emission of young stars have prevented any use of the UV excess
to infer mass accretion rates. In this paper we
model this excess emission and determine the accretion rates for
a number of intermediate mass T Tauri stars for the first time. 

Similarly, there is no way to measure accurate accretion rates for TTS
with very high amounts of excess radiation, which dominate completely over
the emission from the stellar photosphere.  Because the spectral type
of the underlying star cannot be determined accurately, the intrinsic
stellar spectrum is uncertain.  Moreover, because the stellar photospheric
lines can only be detected over a limited range of wavelength, the observed
colors of the stellar photosphere are also poorly known.  This means that
the reddening of the stellar photosphere cannot be determined accurately.
However, extinction corrections are essential for an accurate measurement
of the excess ultraviolet continuum.
In Paper~I, we proposed a method to estimate the extinction and accretion rates
towards the continuum stars by assuming that the spectral energy distribution
is similar to that of the extracted accretion emission of less
veiled TTS. Here we test the validity of this method with measurements extending
to much shorter wavelengths.

In \S 2 we review the shock model, \S 3 outlines the analysis of
archive IUE data, 
in \S 4 we show the agreement of the predicted
shock emission with IUE spectra for low mass T Tauri stars and use the
model to derive accretion rates of continuum
stars and intermediate mass T Tauri stars. Finally, in \S 5, we
discuss some important implications of our determinations.

\section{Shock model}  

Details concerning the shock model calculations are
presented in Paper~I. In summary, the accreting gas impacts the
stellar surface and shocks to a temperature of $\sim 10^6$ K, releasing
the energy as soft X ray radiation. This radiation is absorbed by the
accretion stream above and the stellar photosphere below the shock and
thermalizes, producing optical and UV emission.  The spectral distribution
of this excess emission can be understood as optically thick
emission from the heated photosphere below the shock, appearing mostly
in the Paschen and Brackett continua, and optically thin emission from
the pre-shock and attenuated post-shock regions, becoming important at
wavelengths shorter than the Balmer threshold. In general, roughly 3/4
of the total column luminosity is emitted by the heated atmosphere and
the rest by the post-shock and pre-shock regions.

The accretion
luminosity and rate depend on two parameters: the fractional surface
coverage of the column, $f$, and the energy flux of the accretion
flow, $\curf = 1/2 \, \rho {v_s}^3$, where $\rho$ and $v_s$ are the
density and the inflow velocity of the accretion stream, respectively.
Therefore, given the stellar parameters (mass, radius, and effective
temperature) the accretion rate for a star is given by fitting $f$ and
$\curf$ to the excess emission spectrum.

In Paper I, we compared the continuum emission predicted by the model
with the optical excess emission of T Tauri stars, extracted from
spectra collected in the wavelength region 3200 \AA\ to 5400 \AA\ (GHBC). In
this wavelength region the shock emission is dominated by the diffuse
emission from the pre-shock plus the emission from the heated
atmosphere. However, a deveiling procedure is required to
separate the excess emission from the photosphere, which introduces
uncertainties due to the reddening correction, selection of template,
etc. In the ultraviolet, on the other hand, shock emission dominates
over the photosphere, allowing a more direct comparison between
observations and shock model predictions. 

\section{Observational data}
\label{sec_obs}

We have extracted ultraviolet fluxes for T Tauri stars from the IUE
Archive for the LWP and LWR (LW) and the SWP (SW) modes of the
spectrograph on board the IUE satellite. 
For the stars analyzed in this paper, we took all the spectra
available in the Archives.

To test the shock model, we combine 
optical and UV
data for the stars, and determine if the model can explain
both the UV emission and the properties of the excess
emission in the optical, in particular, the observed veiling.
However, the intrinsic variability of TTS makes it difficult
to compare optical, LW, and SW spectra for a given star, 
all obtained at different times. Since more than one
LW and SW spectra exist for many TTS, we decided to
calculate a mean spectrum to compare with 
the optical data. When sufficient
IUE observations exist, the spread of the individual
spectra around the mean would give a good indication
of the range  of variability of the UV spectra
for a given star. Also in this case, it would
be reasonable to expect
that the mean level of the optical data at 3200 {\AA}
should fall within the range of the mean levels
of the LW fluxes at the same wavelength. Unfortunately,
we do not have a similar constraint
for the SW spectra,
since the ranges in which
the SW and the LW spectra
have
better signal to noise ratio (S/N) do not overlap.
We restricted our test
to the demonstration
that the model that fits both the
optical and the LW data fell within
the range of variability of the SW fluxes.
A better test 
will be done when simultaneous UV coverage from short
to long wavelengths is obtained; our forthcoming
{\it HST/STIS} observations will provide such data.

To calculate the mean spectra for a set of 
LW (or SW) spectra of a given star, we carried out the
following analysis.
We fitted each 
spectrum, $F_i (\lambda)$, logarithmically as
in log space,
\begin{equation}
\log F_{s,i}(\lambda) = A_i + B_i (\log \lambda - \log \lambda_0),\ \ i=1,...,N,
\label{linea}
\end{equation}
where $N$ is the number of LW or  SW spectra for the star.
The fit was done to the wavelength range 
where the
S/N per resolution element is highest,
2400  - 3100 {\AA} for
the LW spectra, and
1700 - 2000 {\AA} for the SW spectra, with
$\lambda_0 = $ 2700 {\AA} and 1800 {\AA}, respectively.
From
the dispersion of the
data  around the
fitted line, $\sigma_i$, 
we can calculate the uncertainties
in the absolute level $A_i$, and in the slope $B_i$,
$\sigma_{A_i}$ and $\sigma_{B_i}$ (Bevington 1969)
for each spectrum.

In many cases, the noise in the continuum level
of the TTS spectra in the IUE Archive is very large,
because the 
exposure times were only long enough to
obtain the fluxes of the emission lines.
We expect that the uncertainty in the 
continuum level, $\sigma_{A_i}$, gives
a good estimation of this noise,
so we define a mean spectrum as
\begin{equation}
\bar F(\lambda) = \frac{\sum_i {w_i} F_i(\lambda)}{\sum_i {w_i}},
\label{meanspec}
\end{equation}
with $w_i = 1/{\sigma_{A_i}}^2$.

We can also calculate weighted means for the slope and the absolute flux
\begin{equation}
\bar A = \frac{\sum_i w_iA_i}{\sum_i w_i},
\end{equation}
and
\begin{equation}
\bar B = \frac{\sum_i {w_i}^{'} B_i}{\sum_i {w_i}^{'}}
\end{equation}
where ${w_i}^{'} = 1/\sigma_{B_i}^2$,
and the dispersion of the slope and the flux
level of all spectra around these weighted means,
\begin{equation} 
{\sigma_{\bar A}}^2 = \frac{\sum_i (A_i - \bar A)^2}{(N-1)}, 
\label{sigmaabar}
\end{equation}
and
\begin{equation}
{\sigma_{\bar B}}^2 = \frac{\sum_i (B_i - \bar B)^2}{(N-1)},
\label{sigmabbar}
\end{equation}
These dispersions give a measure of the intrinsic variability of the UV
spectra of each star.
Although the above analysis is strictly valid only if the scatter in the
flux level of the spectra is normally distributed, which is not formally
satisfied for all spectra, it provides an acceptable
method to measure the characteristics of the 
spectra and their variations.

\section{Results}

\subsection{Testing the model with a low mass T Tauri star: BP Tau}

\begin{figure}
\plotone{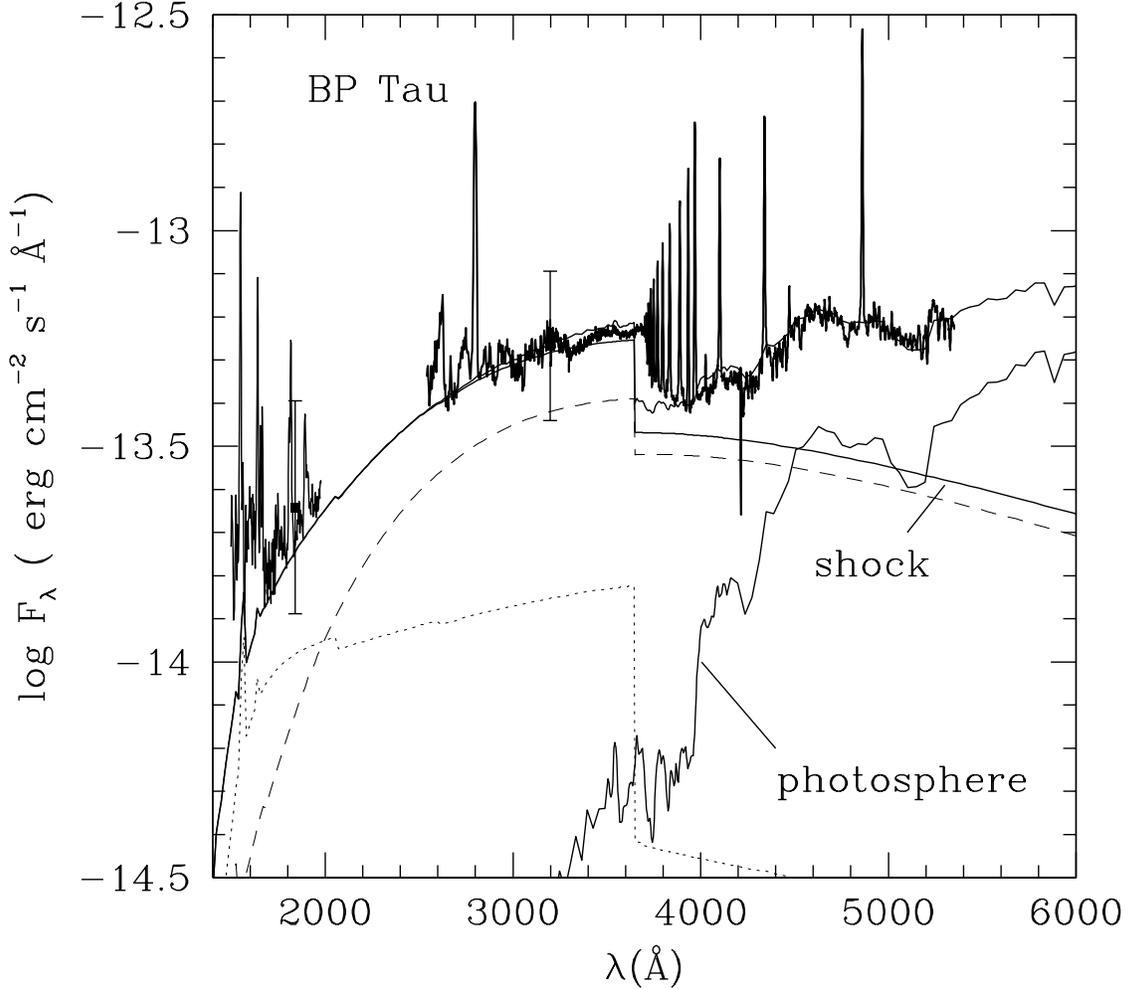}
\caption{Comparison between observed fluxes and shock
model emission for BP Tau.
Observations (thick solid line) are: optical
spectrum from GHBC, mean fluxes LW and SW IUE fluxes,
defined by equation (\ref{meanspec}). 
IUE fluxes have smoothened by 5 pixels, for better visualization.
Mean flux
level at 2700 {\AA} and 1800 {\AA} and range
of variability are indicated by the squares and error bars.
The theoretical model
(thin solid line) is composed of
the emission from the shock and from the stellar photosphere.
The shock emission, in turn, is the sum
of the emission from attenuated post-shock and
diffuse
pre-shock (dotted line) and the heated atmosphere (dashed
line) (Paper I).The shock parameters used are: log $\curf$ = 11.5,
$f$ = 0.007, and $A_V$ = 0.51}
\label{bptau}
\end{figure}

In Paper~I we showed how the accretion shock model
simultaneously
explained the slope of the Paschen and the Balmer continua by the
combined emission from optically thin and thick regions.
The UV fluxes provide a longer ``level arm'' in wavelength
to test the predictions far in the Balmer continuum,
which was traced only over a few hundred {\AA}
in the optical. For a direct application of the shock model,
we have compared theoretical spectra with
the combined optical, LW, and SW data
of the low mass T Tauri star BP Tauri,
all dereddened according
to the values for $A_V$ from GHBC.
Figure \ref{bptau} shows  this comparison.
The optical fluxes are taken from GHBC.
The UV spectra are weighted means,
calculated as described in \S \ref{sec_obs},
of 60 LW spectra and 16 SW spectra of varying S/N levels.
The value of $\bar A$ and error bars corresponding to 1 $\sigma_{\bar A}$
at $\lambda = $ 2700 {\AA} and 1800 {\AA}
are also shown.  
The LW data has been scaled by
a factor 0.9 to match the optical data, which is well
within the 1 sigma of the variability of the flux.

The total model emission is the sum of
the stellar photosphere
and the shock,  which in turn is the sum of the emission
of the heated photosphere and the
pre-shock emission (Paper I.)
The parameters of the shock shown in Figure \ref{bptau} are the same
as used in Paper I, while the  stellar photosphere was taken from the
Bruzual \& Charlot (1993) spectral library.
The model emission can explain the observed
optical and UV fluxes very well, both in slope and in absolute
level, within the range due to variability.
In addition, the model yields a
veiling at 4900 {\AA} of 0.8, consistent with the
measured value of 0.7 (GHBC).
The very good agreement between theory and observations
renders strong support for the
physical parameters derived in Paper I and for the
accretion shock model in general.
In addition, since the UV emission is easily extincted (with
a $A_{2400}/A_V = 2.5$ for a normal interstellar
extinction curve [Mathis et al. 1990]) the agreement 
makes firmer the extinction determination from
the optical spectrum in GHBC.

\subsection{The continuum stars}
A number of TTS show a very high amount of excess radiation that
completely dominates over the emission from the stellar
photosphere. This makes it very difficult to estimate 
the spectral type and color of
the underlying star, and renders estimates of the amount of
interstellar reddening very uncertain. In Paper~I, we
proposed a method to estimate the extinction 
towards the continuum
stars by assuming its spectral energy distribution to be similar to that
of the extracted accretion emission of less veiled TTS,
at least for $\lambda < $ 5400 {\AA} (the
upper wavelength limit of our spectra.) The strong
extinction dependence of the UV spectra 
and the larger wavelength coverage of the combined
optical and ultraviolet spectra offer an efficient way to test
this method and refine our extinction determination. 

\begin{figure}
\plotone{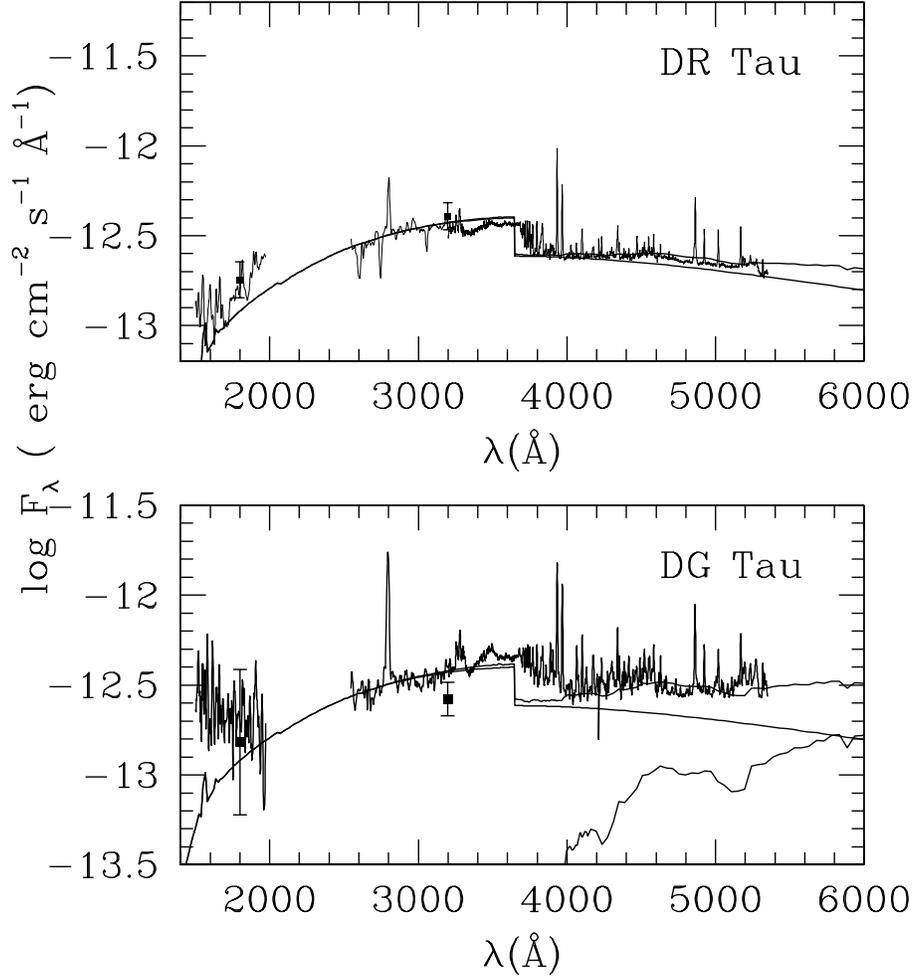}
\caption{Observed (thick line) and calculated (thin line) spectral energy 
distribution for the continuum stars DR Tau and DG Tau. The observations have been de-reddened 
and the shock models calculated for the parameters given in Table 1.
The IUE LW spectra have been scaled to match the optical
spectrum at 3200 {\AA}. The actual
mean flux
levels at 2700 {\AA} and 1800 {\AA} and ranges
of variability are indicated by the squares and error bars.
}
\label{drdg}
\end{figure}

In Figure \ref{drdg} we show optical spectra from GHBC and 
weighted mean SW and LW
spectra for DG Tau and DR Tau, both highly
veiled stars (Basri \& Batalha 1990;
Hartigan, Edwards,
\& Ghandour 1995; Hessman \& Guenther 1997).
The values of $\bar A$ and $\sigma_{\bar A}$ are also
shown.
DR Tau has 31 spectra in the IUE Archive, 
20 LW and 11 SW. 
However, we found that a significant number of them,
9 LW and 5 SW,
were obtained during the period Jan 20-25, 1989.
We used these spectra in our analysis 
to mitigate the uncertainties introduced by variability.

We have adjusted the reddening to give a better fit
to the optical and LW data, the latter scaled to the optical,
subject to the condition that the model should
be within the variability range of the SW spectra.
The scalings of the LW fluxes to match 
the optical at 3200 {\AA} were 0.9 for DR Tau and 1.3 for DG Tau, both
within the range expected from variability in the LW spectra. The 
values of the reddening and of the shock model parameters
are given in Table 1;
the new values are consistent with those obtained in 
Paper I. 

We have also estimated the mass accretion rate
using the expression
(Paper~I):
\begin{equation}
\mdot =  \frac{{8 \pi R}^2}{{v_s}^2} \, \curf f,
\label {Mdot}
\end{equation}
assuming $M_* = 0.5 \msun$ and $R_* = 2 \rsun$.
These values are also shown in Table 1.
We stress again the point made in Paper I,
namely that the continuum stars have accretion columns
with similar values of energy flux as the more typical
TTS, but their accretion columns cover a larger fraction
of the stellar surface, resulting in higher mass
accretion rates. In
support of our conclusion, Ardila
and Basri (2000) find a correlation between  
the mass accretion rate and the accretion shock 
surface filling factor
in their study of the variability of
the IUE spectra of less veiled T Tauri stars.

\subsection{Intermediate mass T Tauri stars}
For intermediate mass TTS with early spectral types (G to early K),
the stellar emission dominates over the accretion
emission in the optical. This makes it difficult to
measure the amount of accretion luminosity, and therefore the
accretion rate for these stars, by conventional methods such as
veiling measurements. This problem is relaxed in the
UV spectral region, where the stellar contribution is lower and
therefore the relative contribution of the excess emission higher.
Measurement of the excess energy above the photosphere yields
the most direct and reliable determination of the accretion luminosity and
thus of the mass accretion rate. 

We have applied the shock model to explain the dereddened fluxes
of a sample of
intermediate mass TTS: SU Aur, GW Ori, and T Tau N.
These stars are generally brighter in the UV and the S/N
is somewhat higher than for the later type TTS.
Adopted stellar parameters, 
like spectral type, radius, and distance
were taken from Kenyon \& Hartmann (1995) (T Tau, SU Aur),
and Cohen \& Kuhi (1979) (GW Ori) 
and are shown in Table 2.
We calculated shock models
for the range of masses and 
radii covered for these stars: 2.2 $\msun$, 3.2 $\rsun$,
for T Tau and SU Aur;  
3.5 $\msun$, 8 $\rsun$ for GW Ori.

An advantage of the earlier type TTS, as compared to the typical
TTS, is that estimates of the interstellar extinction can be made with
less ambiguity, since the observed optical color (V-R) is
expected to be less contaminated from shock emission.
Nonetheless, since there is a range
of extinctions available in the literature, we checked
all the extinctions of the stars in the sample in a consistent
way, adopting standard colors for the adopted spectral
types from Kenyon \& Hartmann (1995)
and reddening law from Mathis (1990).
The adopted extinctions are shown in Table 2. 

\begin{figure}
\plotone{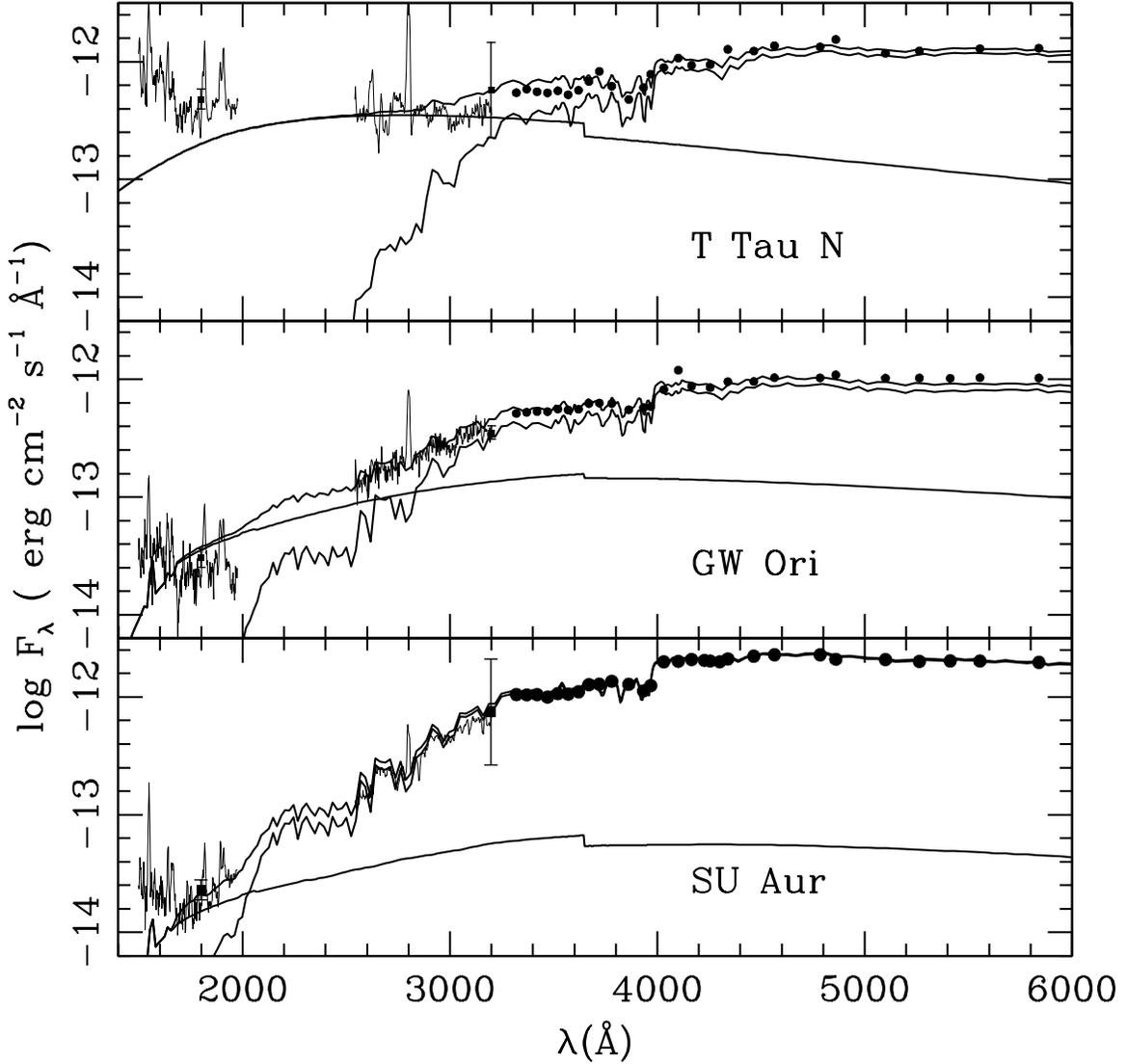}
\caption{De-reddened 
observed (heavy solid line and circles) and 
calculated (light solid line) spectral energy distribution for the 
%intermediate mass T Tauri stars DI Cep, T Tau N, GW Ori, and SU Aur. 
intermediate mass T Tauri stars T Tau N, GW Ori, and SU Aur.
Optical fluxes correspond to the narrow-band
photometry of Kuhi (1974). 
Mean flux 
level at 2700 {\AA} and 1800 {\AA} and range 
of variability are indicated by the squares and error bars.
The contributions from the shock
and the photosphere are shown for reference (cf. Figure \ref{bptau})}
\label{gktts}
\end{figure}

In Figure \ref{gktts}, we show the dereddened IUE and
optical data for the stars. 
The IUE fluxes are weighted means obtained as described in \S \ref{sec_obs}.
Optical fluxes are from
the blue narrow band photometry of Kuhi (1974).
The stellar photosphere is taken from the Bruzual-Charlot (1993)
library. 

The models that provide the best fit to the combined
optical and UV data are shown in Figure \ref{gktts}.
In addition to fitting the SED, we have required that
the model yield a veiling consistent with observations.
The predicted veiling at 6000 {\AA} are shown in Table 2.
The models for T Tau and SU Aur
predict essentially no optical veiling, $\le 0.1$, as observed (Basri \& Batalha 1990).
For GW Ori, the model predicts a veiling of 0.15,
consistent with the veiling at H$\alpha$ of 0.1-0.2
found by Basri \& Batalha (1990).

By fitting the observed SED
with the shock model plus stellar atmosphere we
obtained values for both the filling factor and the energy
flux. These values, together with the stellar parameters, enabled us to
estimate the accretion rate for the stars from equation 
(\ref{Mdot}).
The accretion rate values are presented in Table~2. SU Aur and T Tau
show values of $\mdot$ which are only slightly higher than the
mean of the K7-M3 TTS
(GHBC). In contrast,  
we find a higher value for the mass accretion rate
for GW Ori,  comparable to those
inferred for the low
mass continuum stars
(Table 1). The star is not so highly veiled
as the continuum stars because of its brighter photosphere,
but otherwise it seems to be similar to their low mass
counterparts in that the accretion column carry similar
energy flux, log $\curf \sim 11 - 12$, but the
surface coverage of the accretion columns is larger.

\section{Discussion}

\subsection{Calibration of IR indicators of $\mdot$}

The UV derived measurements of the accretion luminosity in
intermediate mass TTS also serve as a test of other accretion
indicators, namely infrared emission lines.  Such diagnostics are
ultimately the most desirable for intermediate mass stars since they
can be observed in large samples of objects much more easily than UV
continuum fluxes.  In order to apply the line indicators, however,
the line luminosity - accretion luminosity relation must first be
calibrated with blue/UV continuum data.

Muzerolle, Hartmann, \& Calvet (1998b) observed the infrared emission lines Pa$\beta$
and Br$\gamma$ for the GHBC sample of low mass TTS
in Taurus, in addition to a separate sample of intermediate mass TTS
in both Taurus and Orion.  These observations were taken with CRSP on
the 2.1m at KPNO in January 1998 (see Muzerolle et al. 1998b for further
details of the observations, data, and analysis of the low mass star
sample).  A tight correlation between the accretion luminosity
(as determined from the blue excess in GHBC) and
the Pa$\beta$ and Br$\gamma$ line luminosities for the low mass TTS
was found.  This correlation can be employed as a calibration
for determining accretion luminosities from the line luminosities -
important for higher mass and highly extincted young objects where
the short wavelength continuum is difficult or impossible to observe.

\begin{figure}
\plotone{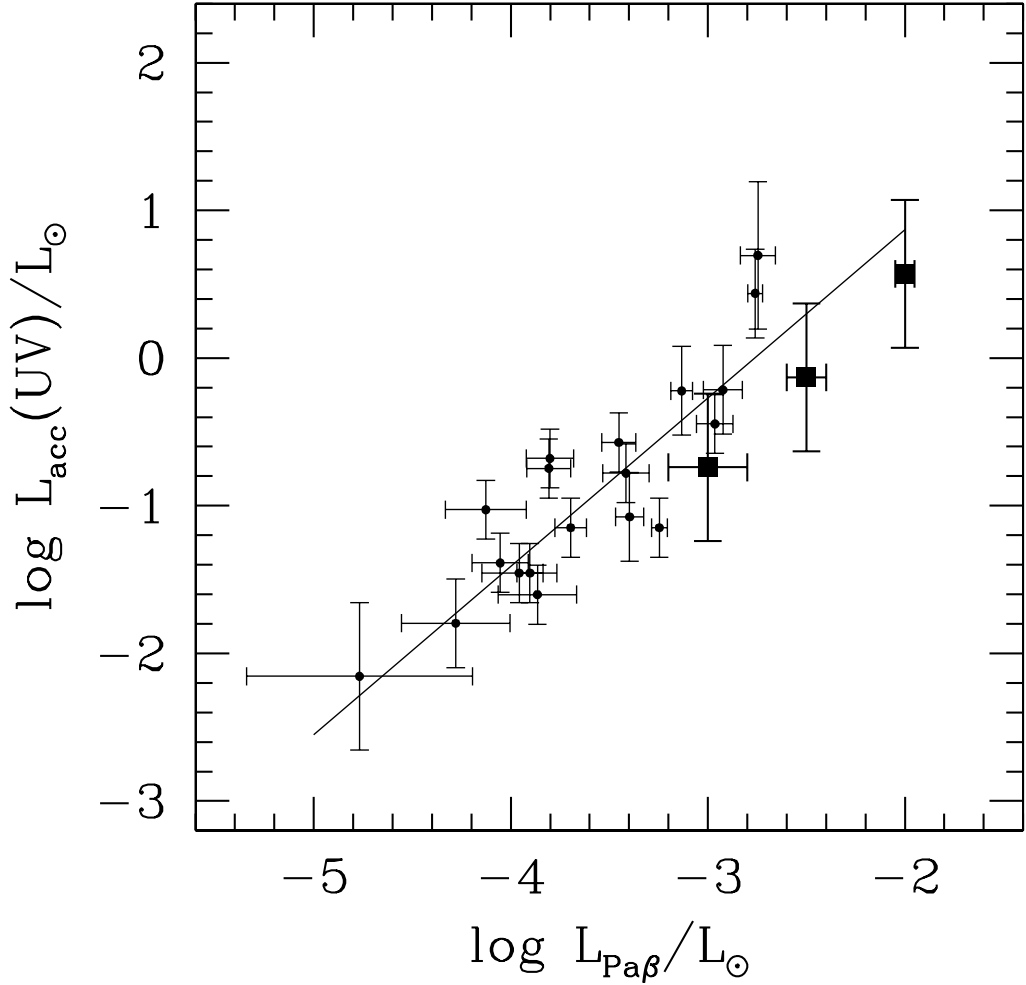}
\caption{The distribution of Pa$\beta$ emission line
luminosities vs. UV continuum excess-derived accretion luminosities,
for low-mass TTS (small dots) and intermediate-mass TTS (large squares).
The best-fit line to the low-mass TTS data is also shown for reference.
Note how the intermediate-mass stars roughly follow this trend.}
\label{figpab} 
\end{figure}

We plot the Pa$\beta$ line luminosities vs. UV derived accretion
luminosities of the intermediate mass stars considered in this
paper in Figure \ref{figpab}, on top of the
low mass TTS sample.  The three stars follow the trend seen in the
lower mass stars, which supports the extension of the
emission line calibration to higher mass stars, although a larger
sample is needed to make a definitive test.
Assuming this correlation holds regardless of stellar mass, we 
can derive {\it independent} estimates of the accretion luminosity in
the three stars from the line emission.  The resulting accretion rates
are shown in Table 2, and are in excellent agreement with the values derived
from the UV excess.

\subsection{Accretion and the GW Ori disk ``gap''}

We derive a high accretion rate for GW Ori, $\sim 4 \times 10^{-7} \msunyr$,
compared with typical lower mass T Tauri stars.  Unless GW Ori is undergoing
a current, rare epoch of rapid accretion, this requires a relatively
large disk mass reservoir.  Consistent with this finding, the large
submm dust continuum flux observed from GW Ori suggests a disk mass
of $\ge 0.3 \msun$ within a radius of 500 AU (Mathieu \etal 1995).
With such a large disk mass, the GW Ori accretion rate 
can be maintained in principle for $\ge 7.5 \times 10^5$~yr, approaching
the estimated stellar age of $\sim 10^6$~yr (Mathieu, Adams, \& Latham 1991).
However, GW Ori is also a spectroscopic binary, with a period
of 242 days (Mathieu \etal 1991); its bimodal spectral 
energy distribution (SED) has been interpreted as requiring
a disk gap between $\sim 0.17$ and 3.3 AU, 
evacuated by the companion.  Because the gap can be filled in with
relatively small amounts of dusty material (cf. the case of DQ Tau;
Mathieu \etal 1997), this interpretation of the SED
makes it highly unlikely that material from the outer disk is accreting
across the gap.  Our accretion rate, coupled with a reasonable system
lifetime, would then require an uncomfortably large disk mass within $\sim 0.2$~AU.

Alternatively, the long wavelength IR excess could be dominated by a dusty
envelope, not the disk.  Calvet \etal (1994) showed that an infalling dusty
envelope with an outflow cavity could produce a double-peaked SED qualitatively like
GW Ori's if viewed roughly along the outflow axis.  In this model, the star and
inner disk is essentially not extincted, accounting for the optical to near-IR emission,
while the far-IR peak at $\sim 30 \mu$m is produced by the dusty envelope. 
The near-IR emission of the envelope is reduced or eliminated by the evacuation 
at small radii, both by the outflow and by rotation 
(the latter causes material to fall onto the disk rather than into the central star).
Thus, in the envelope model the ``gap'' apparent in the SED 
is a hole in the infalling envelope, not in the disk.  

By eliminating the large disk gap, the envelope model makes it easier to 
suppose that disk material can accrete past the binary, as in DQ Tau,
and maintain the accretion rate we determine for a reasonable lifetime.

\section{Summary}

We have extended the accretion shock models of Calvet \& Gullbring (1998)
to shorter wavelengths for comparison with ultraviolet data of
accreting T Tauri stars. 
The shock models agree well with the observed spectral energy
distribution, helping to support out previous determinations of
accretion rates, and providing further evidence in favor of
our methods of determining mass accretion rates and extinctions in continuum
stars.  We provide the first direct estimates of accretion rates 
for T Tauri stars of early spectral types (G2 $-$
K0). These estimates agree with accretion rates estimated 
from the near infrared hydrogen line strengths (Muzerolle et
al. 1998b).  We confirm the result of Paper I that the
continuum stars have similar energy fluxes as less veiled
stars, but larger surface coverage.

We thank Chris Johns-Krull and Jeff Valenti for
providing data for early versions of this paper.
The ultraviolet data used in this paper were obtained from the
Multimission Archive at the Space Telescope Science Institute (MAST).
STScI is operated by the Association of Universities for Research
in Astronomy, Inc., under NASA contract NAS5-26555. Support for
MAST for non-HST data is provided by the NASA Office of
Space Science via grant NAG5-7584, and by other grants and contracts.
This work was supported in part by a grant
from the Swedish Natural Research Council, and
by NASA grant NAG5-4282 and
NASA through grant number GO-08317.01-97A from the Space Telescope
Science Institute.

\clearpage
\begin{center}
\begin{table}[t]
\begin{tabular}[h]{lcccc}
\tableline
\multicolumn{5}{c}{\bf TABLE 1} \\
\multicolumn{5}{c}{\bf SHOCK PARAMETERS FOR ``CONTINUUM''  }\\
\multicolumn{5}{c}{\bf T TAURI STARS }\\
\multicolumn{5}{c}{}\\
\tableline
Object & $A_V$ & log $\curf$& $f$& $\mdot$\tablenotemark{a}\\
       &       &                & & $\times 10^{-7}\msunyr$ \\
\tableline
DR Tau & 1.2 &  11.5 & 0.05 & 3 \\
DG Tau & 1.6 & 11.5 & 0.05 & 5  \\
\tablenotetext{a}{$M = 0.5 \msun$
and $R = 2 \rsun$ have been assumed.}
\end{tabular}
\end{table}
\end{center}

\clearpage
\begin{center}
\begin{table}[t]
\begin{tabular}[h]{lccccccccc}
\tableline
\multicolumn{10}{c}{\bf TABLE 2} \\
\multicolumn{10}{c}{\bf PARAMETERS FOR INTERMEDIATE-MASS T TAURI STARS}\\
\multicolumn{10}{c}{}\\
\tableline
Object & Type & $R_*$ & $A_V$ &  distance & log $\curf$\tablenotemark{a} & $f$ & $r_{6000}$  & $\mdot$ & $\mdot_{\rm{NIR H}}$\\
       &      & $\rsun$ &  &  pc &  & &  & $\times 10^{-7}\msunyr$  & $\times 10^{-7}\msunyr$ \\
\tableline
T Tau N & K0 & 3.7 & 1.7 & 140 & 12 & 0.003 & 0 &0.4 & 0.9 \\
SU Aur  & G2 & 3.2 & 0.9 & 140 & 11 & 0.01 &0 & 0.1 & 0.1 \\
GW Ori & G5 & 8.3 & 0.8 & 440 & 11 & 0.03 & 0.15 & 4 & 3 \\
\tableline
%\tablenotetext{a}{Values binned to the model grid in Paper~I}
\end{tabular}
\end{table}
\end{center}

\clearpage

\clearpage

\clearpage

\clearpage

\end{document}